\documentclass[journal]{IEEEtran}
\usepackage{blindtext}
\usepackage{graphicx,subfig}
\usepackage{lipsum}
\usepackage{tabularx}

\usepackage{algcompatible}
\usepackage{amsmath}
\usepackage{amssymb}
\usepackage{graphicx,adjustbox}
\usepackage{tikz}

\usepackage[justification=centering]{caption}
\usepackage{algorithm}% http://ctan.org/pkg/algorithms
\usepackage{algpseudocode}
\usepackage[T1]{fontenc}% optional T1 font encoding
\usepackage{amsmath}
\usepackage{tasks}
\usepackage{url}
\usepackage{color}
\usepackage{diagbox}
\newtheorem{example}{Example}
\newtheorem{lemma}{Lemma}

\newtheorem{theorem}{Theorem}
\newtheorem{remark}{Remark}

 \usepackage{tikz}
\usetikzlibrary{shapes,backgrounds}
\usetikzlibrary{arrows,chains,matrix,positioning,scopes}
\usetikzlibrary{arrows.meta}
\usetikzlibrary{calc}
\usetikzlibrary{arrows,decorations.markings}
\newcommand{\MR}{M^{\rightarrow}}
\newcommand{\ML}{M^{\leftarrow}}
\newcommand{\V}{{\cal V}}
\newcommand{\Source}{{\cal S}}
\newcommand{\D}{{\cal D}}
 	\definecolor{darkgreen}{rgb}{0.0, 0.5, 0.0}
 	\definecolor{rose}{rgb}{1.0, 0.01, 0.24}
 	\definecolor{move}{rgb}{0.76, 0.13, 0.28}
\begin{document}
\title{Routing and Network Coding over a Cyclic Network for Online Video Gaming}
%\author{Michael~Shell,~\IEEEmembership{Member,~IEEE,}
%        John~Doe,~\IEEEmembership{Fellow,~OSA,}
%        and~Jane~Doe,~\IEEEmembership{Life~Fellow,~IEEE}
%\author{Author 1, Author 2, Author 3, Author 4}
\author{
  Marwa Dammak, ETIS-UMR8051, ENSEA/ University of Cergy Pontoise/ CNRS, France \\and LETI, ENIS/ University of Sfax, Tunisia

   Iryna Andriyanova, ETIS-UMR8051, ENSEA/ University of Cergy Pontoise/ CNRS, France 

  Yassine Boujelben, NTSCOM, ENET'COM/ University of Sfax, Tunisia

   Noura Sellami, LETI, ENIS/ University of Sfax, Tunisia 
}
%\thanks{M. Shell was with the Department
%of Electrical and Computer Engineering, Georgia Institute of Technology, Atlanta,
\algnewcommand\algorithmicswitch{\textbf{switch}}
\algnewcommand\algorithmiccase{\textbf{case}}
\algnewcommand\algorithmicassert{\texttt{assert}}
\algnewcommand\Assert[1]{\State \algorithmicassert(#1)}%
% New "environments"
\algdef{SE}[SWITCH]{Switch}{EndSwitch}[1]{\algorithmicswitch\ #1\ \algorithmicdo}   {\algorithmicend\ \algorithmicswitch}%
\algdef{SE}[CASE]{Case}{EndCase}[1]{\algorithmiccase\ #1}{\algorithmicend\     \algorithmiccase}%
\algtext*{EndSwitch}%
\algtext*{EndCase}%
%\markboth{IEEE Communication Letters,~Vol.~XX, No.~X, XXXX~201X}%
%{Shell \MakeLowercase{\textit{et al.}}: Bare Demo of IEEEtran.cls for IEEE Communications Society Journals}

\newcommand*\xor{\mathbin{\oplus}}
\newcommand{\ndiv}{\hspace{-4pt}\not|\hspace{2pt}}

\maketitle

\begin{abstract} 
Online video games are getting more popular, attracting a continuously growing number of players. 
The main performance metrics of this application on the network level are packet ordering, communication throughput and transmission latency. 
Nowadays there is an interest in cyclic logical network topologies for online gaming, due to the easiness to preserve the packet order over cycles.
Unfortunately, this approach increases the end-to-end transmission delays.  
In our paper, two main contributions are therefore presented. 
Firstly, it is shown that one can improve the latency of a gaming protocol over a single-cycle topology by the network coding (NC) approach. The corresponding NC-based routing protocol has been designed and analysed; it outperforms the best routing protocol without NC.   
Secondly, from the NC viewpoint, the example of online gaming is not a trivial one, given that the corresponding communication protocol is not multicast. Therefore, it is shown that there exists a NC gain (namely, up to 14\% in transmission latency) even in case of a mixed communication protocol with broadcast and unicast transmission flows.  

\end{abstract}


%%%%%%%%%%%
\section{Introduction}
\label{sec:intro}
%%%%%%%%%%%

\IEEEPARstart{T}o improve the performance of emerging real-time group-based multiuser applications, e.g., multiplayer online games, Internet-of-Things or mobile group services, a distributed 2-tier architecture has been largely proposed  \cite{Barri2016}-- \cite{NaorDas2017}. 
In such an architecture, user instances are distributed among multiple servers in order to meet the delay constraint imposed by the real-time service requirement. Thus one server handles one communication group, usually by organizing a multicast transmission among the nodes.
Basically, there are two topologies to communicate within a communication group: trees and cycles. 
A cycle-based network is well suited for applications that require ordering, low-overhead control protocol and inherent failure tolerance as illustrated in \cite{Babay}--\cite{ring3}. 
Compared to trees, the cycle topologies improve throughput and reliability, but cause an additional latency. 

The focus of our paper is online video gaming -- a user application of increasing popularity (e.g., the number of active players in $League$ $of$ $Legends$ raised from $11.5$ billions/month in 2011 to more than $100$ billions/month in 2016). It belongs to the class of real-time multiuser applications, and it is evaluated over such performance metrics as {\sl throughput}, {\sl transmission latency} and {\sl packet ordering}, the latter ensuring consistency within the game \cite{ordre2016}.
Given the order-preserving setup of online gaming, it is therefore relevant to perform the routing over a cyclic logical network as it has been already proposed in \cite{yb}\footnote{There, logical cycles are organised to gather players using some proximity critera, e.g., their geographical closeness on the game map \cite{yb}.}. 
However, the issue of transmission latency should be properly addressed.

Our proposition is to improve the latency over cyclic topologies by using Network Coding (NC) based protocols \cite{SolFra}.
It is known that NC improves the transmission delay in multicast \cite{nc2017} and some unicast \cite{unicast} scenarios, and also for several interesting cases over cyclic networks \cite{nc2017}--\cite{3colors}.
However, the specificity of online gaming traffic is to be taken into account -- the communication between a game server and player nodes is not based on multicast but is a mixture of one broadcast and multiple unicasts which differs from the state-of-the-art of NC protocols.
We consider this particular communication scheduling and proceed as follows.
First, an efficient routing protocol (without NC) with small communication latency is built as a reference. 
Then, a NC scheme for online gaming communication protocols over cyclic networks is proposed,  
and its gain in terms of latency is estimated.
For the seek of simplicity, the network with a single cycle is considered, this corresponds to one single proximity-defined logical subnetwork within a game.

%%%%%%%%%%%%
\section{System Model and Notation}
\label{sec:model}
\begin{figure}
	\centering
	\includegraphics[width=0.8\columnwidth]{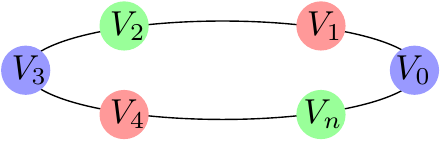}
	\caption{Single-cycle topology for online gaming with $n=5$. Also, $V_0=S$. Node subsets $\V_1$, $\V_2$ and $\V_3$ are given by blue, red and green.}
	\label{ring}
\end{figure}

%%%%%%%%%%%%
Let the logical network of an online game contain a single cycle, connecting the game server $S$ and $n$ game players.  
It is represented by a cycle $G = (V, E)$ (see Fig.\ref{ring}) with the set of nodes $\V=\{V_0=S, V_1, \ldots, V_n\}$, and the set of links $E$. 
For simplicity, let the network be homogenous so that each link $e \in E$ may transmit one packet per time unit (t.u.).
Online game traffic is periodic with {\it a communication period} $T$ [t.u.] and contains $n+1$ flows as follows: 
\begin{itemize}
\item a broadcast flow from $V_0$ to all $V_i  \in V \backslash V_0$ to communicate the current game instance;
\item  $n$ unicast flows from any $V_i  \in V \backslash V_0$ to $V_0$ carrying players' actions. 
\end{itemize}
Also, let us denote by $L$ the total number of packets transmitted through the network during one {\it communication period} $T$.

The directions of $n+1$ flows described above imply two source-destination sets:
\begin{itemize}
\item $\Source_1=\{V_0\}$ and $\D_1 =\{V_1, \ldots, V_n\}$: the node $V_0$ broadcasts a common message $M_0$  to the set $\D_1$; 
\item $\Source_2=\{V_1, \ldots, V_n\}$ and $\D_2 =\{V_0\}$: each node $V_i$ in $S_2$ has a private message $M_i$ to send to $V_0$.
\end{itemize}
Given $n$, $\Source_1$, $\D_1$, $\Source_2$ and $\D_2$ defined above, we distinguish
the {\it routing problem} which aims to find a routing $\cal R$ minimizing the couple $(T, L)$, and
the {\it network coding (NC) problem} which consists to find a network code $\cal C$ and a routing protocol $\cal R_C$ minimizing $(T,L)$.

For later use, let $D \triangleq \lceil n/2 \rceil$ and $d\triangleq \lfloor D/2\rfloor = \lfloor \frac{n+1}{4} \rfloor$.

\subsection{Transmission Rules and Related Transmission Scheduling}
In order to solve the routing and NC problems, transmission rules within $G$ are to be fixed. Let assume them as following:
\begin{enumerate}
\item A node $V_i$, when sending a message $M$, broadcasts it to its closest neighbours $V_{i-1}$ and $V_{i+1}$, where the index addition/subtraction is performed modulo $n+1$ ({\it possibility to broadcast});
\item During one time unit, each node is either in the receiving or in the sending mode ({\it half-duplex regime});
\item A node cannot receive two simultaneous messages from its neighbours ({\it collision}).
\end{enumerate}

For the rules above, it can be shown that the best transmission schedule, giving the smallest $T$ with the best collision avoidance, is determined as follows \cite{3colors}.
If $n+1$ is divisible by 3, i.e. $(n+1)|3$,
then the set of nodes $V$ is divided into $3$ disjoint subsets $\V_1$, $\V_2$ and $\V_3$ so that a node from a subset $\V_i$ is at least at distance $3$ from another node from the same subset\footnote{By distance one understands the number of hops between the two nodes.} (refer to Fig \ref{ring} for illustration).
Further, the {\it 3-phase transmission schedule} is adopted: the data transmission is organised in {\it rounds}, each lasting $3$ t.u.; a subset $\V_i$ is allowed to broadcast during one time unit slot, while other subsets will be only listening,  thus avoiding collisions. 
W.l.o.g., we assume that the subset $\V_i$ broadcasts during the time unit slot $i$, $i=1,2,3$.
When $(n+1)\ndiv 3$, 4 disjoint groups $\V_1, \ldots, \V_4$ are formed under the same distance constraint, and a {\it 4-phase transmission schedule} is put on place.  

%%%%%%%%%%%%%%%%%%%
\section{A Motivating Multicast Example}
%%%%%%%%%%%%%%%%%%%
\label{sec:example}
The state-of-the-art of network-coded communication protocols \textcolor{black}{over cyclic topologies} is mostly based on the multicast scenario, 
where each node $V_i$ has a message $M_i$ to send to all other nodes in the network, for all possible values of $i$.
In this case the use of NC is beneficial. Let us illustrate it on the following example by comparing two communication protocols, with and without NC. 

 \begin{example} \label{ex:1}
Consider the {\it circular routing} for multicast over a single-cycle network with $n+1$ nodes $V_0, \ldots, V_n$. Here the messages are forwarded over the cycle in one direction (clockwise or counterwise), following the {3-phase} or 4-phase transmission schedule described above. 
The transmission continues until any message $M_i$ reaches all the nodes in $V\backslash V_i$.
\end{example}
 
{\lemma \label{lemma:circular-routing}
The minimum communication period $T$ for Example \ref{ex:1} is bounded as
$3n \le T \le 4n$, while the total number $L$ of messages to transmit over the network is $L \le \lfloor (n+1)/3\rfloor T$.
}
\begin{proof}
The proof is straightforward.
If $(n+1)|3$, then exactly $n$ rounds of the 3-phase transmission are needed so that $M_i$ reaches all its destinations.
Otherwise, $n$ rounds of the 4-phase transmission will be used.
As for the result on $L$, note that at most $\lfloor (n+1)/3\rfloor$ messages are sent at each t.u. 
\end{proof}

\begin{example}
\label{ex:2}
Consider the multicast problem over the single-cycle network, where the nodes are allowed to perform NC operations. Then, Algorithm \ref{alg:multicast} below can be adopted. Note that the algorithm was first described in  \cite{3colors}  and here we give it when $(n+1) | 3$ only, for the seek of simplicity.
\begin{algorithm}
\caption{\label{alg:multicast} Multicast with NC \cite{3colors} when $(n+1)|3$}
{\bf Initialisation:}
$V_i$ is allocated to the subset $\V_j$, $j=(i\mod 3)+1$, $0 \le i \le n$.
Each $V_i$ has a message $M_i$ to multicast. 

{\bf For $0 \le t\le \lceil n/2 \rceil $, perform the 3-phase transmission:}

During 3 t.u., a node $V_i$ ($0 \le i \le n$) does the following (the order of operations depends on its subset index $j$):

$\quad$1) Reception of a message from the right $\MR_{i+1}(t)$;

$\quad$2) Reception of a message from the left $\ML_{i-1}(t)$;

$\quad$3) Broadcast of the current message $M_i(t)$, where

$\quad$ $M_i(t) = \begin{cases} M_i, & t=0;\\ \ML_{i-1}(t-1) \oplus \MR_{i+1}(t-1), & t>0.\end{cases}$

\end{algorithm} 

\end{example}

{\lemma \label{lemma:multicast}Given the NC-based multicast protocol from Example \ref{ex:2}, 
the minimum communication period $T_{NC}$ is bounded as
$3 \lceil \frac{n}{2} \rceil \le T_{NC} \le 4 \lceil \frac{n}{2} \rceil,$
and the number of messages $L_{NC} \le \lfloor (n+1)/3 \rfloor T_{NC}$.
}
\begin{proof} 
At round $t=0$, $V_i$ receives $M_{i-1}$ from the left and $M_{i+1}$ from the right\footnote{here and below the index addition/subtraction is performed modulo $(n+1)$}. 
At round $t>0$, $V_i$ possesses already the messages $M_{i-t}$ and $M_{i+t}$.  Moreover it receives $M_{i-1}(t)= M_{i-t-1}\oplus M_{i-t+1}$ and $M_{i+1}(t)=M_{i+t-1} \oplus M_{i+t+1}$. So it decodes $M_{i-t-1}$ and $M_{i+t+1}$ by {XOR-ing}: $M_{i\pm t \pm 1} = M_{i\pm1}(t) \oplus M_{i \pm t}$.
At round $t=\lceil n/2 \rceil$, $V_i$ receives the last missing message from its farest node(s).  
As the duration of the round might be 3 or 4 t.u. (depending on $n$), one gets lower and upper bounds on $T_{NC}$.
The calculation of $L_{NC}$ is straightforward.
\end{proof}

Note that, in the half-duplex multicast example, NC improves $T$ by $1/2$ \cite{3colors}.

%%%%%%%%%%%%%%%%%%%
\section{Protocols for Online Gaming}
%%%%%%%%%%%%%%%%%%%
\label{sec:gaming}
The system model for online gaming application is given in Section \ref{sec:model}.
The sets $\Source_1$, $\D_1$, $\Source_2$ and $\D_2$ are specific to online game protocols and give rise to a non-multicast scenario.
Therefore, even if Algorithm \ref{alg:multicast} could a priori be used for online gaming applications, it does not guarantee minimum values for $T$ and $L$. Moreover, the amount of calculations to perform would be quite high, which is an issue for a real-time, bandwidth-consuming application such as online gaming. 

Let us design appropriate routing and NC-based protocols for the setup in Section \ref{sec:model}.

%%%%%%%%%%%%%%%
%%%%%%%%%%%%%%%
\subsection{Optimized Routing}
We propose Algorithm \ref{alg:shortest-path} as a routing protocol without NC for online gaming. It is in fact an optimised version of the {\it shortest-path routing} algorithm. For simplicity, the algorithm is described when $(n+1)|3$. 
\begin{algorithm}[t]

\caption{\label{alg:shortest-path} Routing protocol when $(n+1)|3$}

{\bf Initialisation}:
$V_i$ is allocated to the subset $\V_j$, $j=i\mod 3$, $0 \le i \le n$.
Each $V_i$ has a message $M_i$ to transmit. 

{\bf 3-phase round for $t=0$:}
A node $V_i$ ($0 \le i \le n$) performs the operations as in Algorithm \ref{alg:multicast} for $t=0$.

{\bf 4-phase round for $1\le t \le d$}:\\
During first 3 t.u., at each time unit $\ell$ ($\ell=1,2,3$) the set $\V_i$ broadcasts while the other sets are silent.
Moreover, 
\begin{itemize}
\item \textcolor{black}{if $V_i \in S_{\rightarrow} = \{V_{0},\ldots, V_{D-t}\}\backslash V_0$, it broadcasts \\$M_i(t)=\MR_{i+1}(t-1)$;}
\item \textcolor{black}{if $V_i \in S_{\leftarrow} = \{V_{n-D+t+1}, \ldots, V_{n+1}=V_0\}\backslash V_0$, it broadcasts $M_i(t)=\ML_{i-1}(t-1)$.}
\end{itemize}
During the last t.u., nodes $V_t$
and  $V_{n+1-t}$ broadcast $M_{0}$.\\
{\bf 3-phase round for $d+1\le t < D$} :\\
During first 3 t.u., at each time unit $\ell$ ($\ell=1,2,3$) the set $\V_i$ broadcasts while the other sets are silent.
Moreover, 
\begin{itemize}
\item \textcolor{black}{if $V_i \in S_{\rightarrow} = \{V_{0},\ldots, V_{D-t}\}\backslash V_0$, it broadcasts \\$M_i(t)=\MR_{i+1}(t-1)$;}
\item \textcolor{black}{if $V_i \in S_{\leftarrow} = \{V_{n-D+t+1},\ldots, V_{n+1}=V_0\}\backslash V_0$, it broadcasts $M_i(t)=\ML_{i-1}(t-1)$;}
\item $V_t$ and $V_{n+1-t}$ broadcast $M_{0}$. 
\end{itemize}
\end{algorithm}
To illustrate Algorithm \ref{alg:shortest-path}, Fig.\ref{fig_sim-routing} presents a time diagram for $n=5$ ($D=3$ and $d=1$). At round $0$, the nodes send their own messages in $3$ t.u. At round $1$, $V_0$ and $V_3$ remain silent, thus $V_1$ and $V_5$ send two messages each, and $V_2$ and $V_4$ send messages of their neigbours. 
During round $2$, the nodes $V_2, \ldots, V_5$ should remain silent except $V_4$ which communicates $M_0$ to $V_3$. $V_1$ forwards $M_3$ to $V_0$.

{\remark If $(n+1) \ndiv 3$, then 4 disjoint groups $\V_1, \ldots, \V_4$ are formed following the rules. 
Let $r= (n+1) \text{ mod } 3$ and let $\V_4=\emptyset$ if $r = 0$, $\V_4=\{V_{ \lceil n/2 \rceil}\}$ if $r = 1$ and 
$\V_4=\{V_{ \lfloor n/2 \rfloor},V_{ \lceil n/2 \rceil}\}$ if $r = 2$.
Then
\begin{align*}
\V_1=& \left( \{V_i: i \text{ mod } 3 = 0,  i \le \lfloor n/2 \rfloor \} \cup \right. \\ & \left. \{V_i: i \text{ mod } 3 = r,  i > \lfloor n/2 \rfloor \} \right)\backslash \V_4\\
\V_2=& \left(\{V_i: i \text{ mod } 3 = 1,  i \le \lfloor n/2 \rfloor \} \cup \right. \\ & \left. \{V_i: i \text{ mod } 3 = (r+1)\text{ mod } 3,  i > \lfloor n/2 \rfloor \}\right) \backslash \V_4\\
\V_3=& \left(\{V_i: i \text{ mod } 3 = 2,  i \le \lfloor n/2 \rfloor \} \cup \right. \\ & \left. \{V_i: i \text{ mod } 3 = (r+2)\text{ mod } 3,  i > \lfloor n/2 \rfloor \}\right) \backslash \V_4
\end{align*}
Also, Algorithm \ref{alg:shortest-path} is modified accordingly: the round $t=0$ lasts 4 t.u. (each node subset broadcasts during 1 t.u. and stays silent during 3 t.u.). Thanks to the careful choice of $\V_4$, the rounds for $t>0$ stay unchanged\footnote{
Note that, if $|\V_4|=2$, it seems to be a collision in transmitting $M_{ \lfloor n/2 \rfloor}$ and $M_{ \lceil n/2 \rceil}$ at $t=0$. However there is no loss for $M_{ \lfloor n/2 \rfloor}$ and $M_{ \lceil n/2 \rceil}$, as they are successfully received by $V_{ \lfloor n/2 \rfloor-1}$ and $V_{ \lceil n/2 \rceil+1}$ respectively.
}.

\begin{figure}[t] 
\centering
\centering
  \subfloat[Round $t=0$. $V_0$ receives $M_1$ and $M_5$. $M_0$ is received by $V_1$ and $V_5$. At this instance all nodes possess their own messages.]{\label{fig:edge-a}\includegraphics[width=1.0\columnwidth]{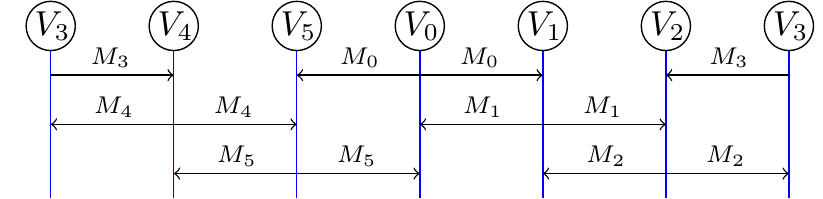}}
  \hspace{5pt}
  \subfloat[Round $t=1=d$. $V_0$ and $V_3$ do not transmit. $V_0$ receives $M_2$ and $M_4$. $M_0$ is received by $V_2$ and $V_4$.]{\label{fig:contour-b}\includegraphics[width=1.0\columnwidth]{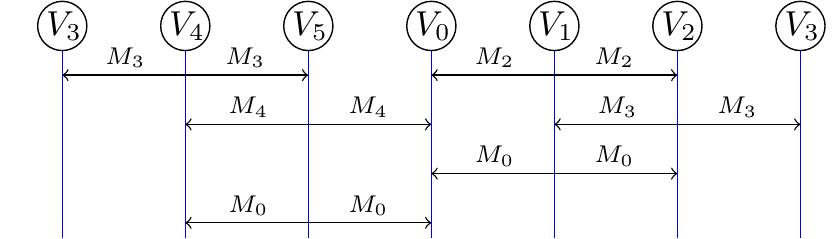}}
  \hspace{5pt}
  \subfloat[Round $t=2$ ($t >d$). $V_0, \ V_2, \ V_3$ and $V_5$ do not transmit. $V_0$ receives $M_3$. $M_0$ is received by $V_3$. End of the communication period $T$.]{\label{fig:contour-c}\includegraphics[width=1.0\columnwidth]{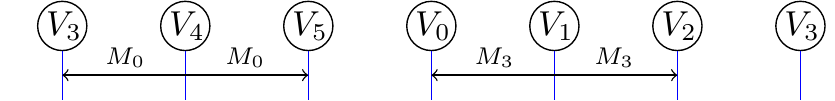}}
  
\caption{Example for $n=5$. Similar to Fig.\ref{ring}, $\V_1=\{V_0, V_3\}$, $\V_2=\{V_1, V_4\}$, $\V_3=\{V_2, V_5\}$.}
\label{fig_sim-routing}
\end{figure}

{\theorem \label{theorem:routing}
The period $T$ of Algorithm \ref{alg:shortest-path} is bounded as 

$ 3 \lceil n/2 \rceil + \lfloor \frac{n+1}{4} \rfloor -2 \le T \le 3 \lceil n/2 \rceil + \lfloor \frac{n+1}{4} \rfloor +1,$  \\
and $L=\lceil n/2 \rceil(\lfloor n/2 \rfloor+3)-1$. 

}

\begin{proof}
The proof is by counting. 
Let $(n+1)|3$.
By shortest-path  routing over the cycle with $n+1$ nodes, a message $M_i$ will be received by a destination in at most $D$ hops.
The hops make part of the rounds of Algorithm \ref{alg:shortest-path}, thus,  for the round $t=0$, $3$ t.u. will be used for one hop.  
The total number of messages sent at $t=0$ is $n+1$.
Moreover, for $1\le t \le d$, we have the following.
The nodes $V_i$ with $1 \le i \le j$ and with $n+1-d\le i \le n$ have two messages to forward: a message $M_j$, $j \in \{i+1,\ldots,i+d\}\cap \{j-d,\ldots,j-1\}$, and $M_0$. These nodes will use one additional t.u. so all of them will last $4$ t.u. Note that the rest of nodes will be silent as they have no new messages to send. 
As for the rounds with $d+1\le t < D$, $M_0$ is now to be transmitted by nodes with indices in $\{d+1,\ldots,\lfloor n/2 \rfloor-1\} \cap \{\lfloor n/2 \rfloor +2, \ldots,n-d\}$. These nodes have no other messages to forward thus they send $M_0$ within $3$ t.u. during which the nodes $V_i$ with $1 \le i \le j$ and with $n+1-d\le i \le n$ forward messages to $V_0$. 
Also, it can be shown that at any round $1 \le t <D$, $n+2-2t$ messages will be sent in total.
Finally, if $(n+1) \ndiv 3$, one more t.u. will be used at $t=0$, and the rest of protocol will be unchanged.
This give us the upper bound on $T$,
$T \le 4d+3(D-d-1)+4$, as well as $L= n+1+\sum_{t=1}^{D-1} (n+2-2t)$.

Also, by better transmission organisation, it is possible to save some t.u. for $t=D-2$ and $t=D-1$, as there are many silent nodes in the network, and the simultaneous transmission by nodes from different subsets does not create collisions. 
It can be shown that this gain is at most 3 t.u.
\end{proof}

%%%%%%%%%%%%%%%
%%%%%%%%%%%%%%%
\subsection{NC-Based Protocol}
Let the nodes perform NC operations. Then the routing protocol above can be modified as follows. 

\begin{algorithm}

\caption{\label{alg:ours}NC-based protocol when $(n+1)|3$}

{\bf Initialisation and round $t=0$:}
As in Algorithm \ref{alg:shortest-path}.

{\bf 3-phase round for $1\le t \le d$:}
At each time unit $\ell$ ($\ell=1,2,3$) the set $\V_i$ broadcasts while the other sets are silent.
\begin{itemize}
\item if $V_i \in S_{\oplus} = \{V_t, V_{n+1-t}\}$, it broadcasts \\ $M_i(t)=\ML_{i-1}(t-1) \oplus \MR_{i+1}(t-1)$;
\item \textcolor{black}{if $V_i \in S_{\rightarrow} = \{V_{0},\ldots, V_{D-t}\}\backslash V_0$, it broadcasts \\$M_i(t)=\MR_{i+1}(t-1)$;}
\item \textcolor{black}{if $V_i \in S_{\leftarrow} = \{V_{n-D+t+1}, \ldots, V_{n+1}=V_0\}\backslash V_0$, it broadcasts $M_i(t)=\ML_{i-1}(t-1)$.}
\end{itemize}
{\bf 3-phase round for $d+1\le t < D$:}
As in Algorithm \ref{alg:shortest-path}.
\end{algorithm}

One can see that the the transmissions at rounds $1\le t \le d$ have been modified.
To illustrate the proposed modification, let us take the example on Fig. \ref{fig_sim-routing}. 
The new protocol would modify the transmission at $t=1$, it becomes as it is shown in Fig.\ref{fig_sim}. Note that 2 t.u. are saved thanks to NC operations.
More generally, the following is stated:

\begin{figure}[t] 
\centering
  \subfloat[Round $t=1=d$. $V_0$ and $V_3$ do not transmit. $V_0$ receives $M_2$ and $M_4$ (encoded). $M_0$ is received by $V_2$ and $V_4$.]{\label{fig:contour-b}\includegraphics[width=1.0\columnwidth]{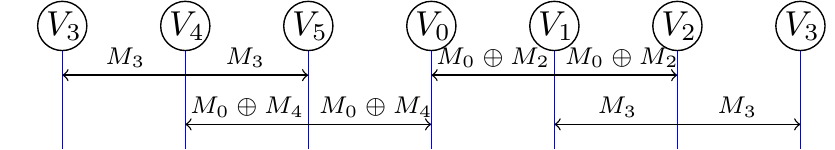}}
\caption{Round $t=1$ of the NC protocol for $n=5$.}
\label{fig_sim}
\end{figure}

{\theorem For Algorithm \ref{alg:ours}, one has
$ 3 \lceil n/2 \rceil  -2 \le T_{NC} \le 3 \lceil n/2 \rceil  +1,$  
and $L_{NC}=\lceil n/2 \rceil(\lfloor n/2 \rfloor+3)-2\lfloor \frac{n+1}{4} \rfloor -1$.
\begin{proof} 
Owing to NC operations, the nodes, having two messages to forward, send their XORs. Note that the nodes receiving XORs are always able to decode new messages.
Thus the rounds with $1\le t \le d$ last $3$ t.u. instead of $4$, and the number of transmitted messages in decreased by $2$ in each round.
By counting the number of messages as transmissions similar as for Theorem \ref{theorem:routing}, one obtains that 
$ 3D/2-3 \le T \le 4 \lfloor D/2 \rfloor +3\lceil D/2 \rceil) $ and $L_{NC} = D(n+3-D)-1-2d$. 
\end{proof}
%%%%%%%%%%%%%%%%%
\section{Discussion}
\label{sec:final}
%%%%%%%%%%%%%%%%%
Table \ref{table} gives the values of $T$ and $L$ for four protocols, described in the paper (multicast circular routing, NC-based multicast from Algorithm \ref{alg:multicast}, optimised protocols without and with NC from Algorithms \ref{alg:shortest-path} and \ref{alg:ours}).
One can see that the optimised routing protocol (Algorithm \ref{alg:shortest-path}) has better performance in terms of $T$ and $L$, compared to multicast protocols. 
Moreover, Algorithm \ref{alg:ours} allows to obtain even larger gains, in particular around 14\% compared to Algorithm \ref{alg:shortest-path} in terms of $T$, when $n$ is sufficiently large.
\begin{table}[t]
\begin{center}
\begin{tabular}{|c|c|c|c|c|}
\hline
$n$& \multicolumn{4}{c|}{$T$ (LB/UB), {\color{blue}$L$}}\\
\cline{2-5}
&Circular routing& Algorithm \ref{alg:multicast}& Algorithm \ref{alg:shortest-path}& Algorithm \ref{alg:ours} \\
\hline 
7&21/28, {\color{blue}42}&12/16, {\color{blue}24}&12/15, {\color{blue}23}&10/13, {\color{blue}19}\\
8&24/32, {\color{blue}72}&12/16, {\color{blue}36}&12/15, {\color{blue}27}&10/13, {\color{blue}23}\\
9&27/36, {\color{blue}81}&15/20, {\color{blue}60}&15/18, {\color{blue}34}&13/16, {\color{blue}30}\\
\hline
\end{tabular}
\caption{\label{table} Comparison  of $T$ (lower bound and upper bound) and $L$ for protocols from Sections \ref{sec:example}-\ref{sec:gaming}.}
\end{center}
\end{table}

Algorithm \ref{alg:ours} is consistent from the game point of view as the arrival order of packets at $V_0$ is predetermined by the placement of nodes within the cycle: 
at each round $t$, $V_0$ receives messages $M_{t+1}$ and $M_{n-t}$. 
This fact can be taken into account during the choice of nodes $V_1,\ldots,V_{n-1}$, in order to design a desired order of message arrivals at $V_0$.%the server node.  

The NC gain of Algorithm \ref{alg:ours} is due to the possibility to broadcast messages to close neighbours (transmission rule 1). 
This condition is easy to satisfy in some kind of networks, i.e., in wireless mesh networks \cite{3colors}.
In wireline networks, broadcast may be implemented by means of the IP-multicast \cite{ip-multicast}.    
Note that, if broadcast is not an option and one has to send messages to the neighbours sequentially (i.e., classical routing in wireline networks), the gain is lost and Algorithm \ref{alg:ours} behaves as Algorithm \ref{alg:shortest-path}. 
Note that, in contrast to the broadcast constraint, the half-duplex constraint (transmission rule 2) does not limit the usefulness of NC. 
For full-duplex transmissions, Algorithms \ref{alg:shortest-path} and \ref{alg:ours} can be modified accordingly. 

Our future work will extend our simple system model to incorporate some properties of gaming protocols, in the order of relevance: a) heterogenous transmission delays between the neighbours in the cycle network; b) difference in the size between the message $M_0$ and other nodes messages and c) packet losses.    
The cyclic topology is also going to be compared with a tree-like topology by simulations in NS-3.

\end{document}